\documentclass[12pt,preprint]{aastex}

\def\eg{{\it e.g., \/}}
\def\ie{{\it i.e., \/}}

\def\gs{\mathrel{\raise0.35ex\hbox{$\scriptstyle >$}\kern-0.6em
\lower0.40ex\hbox{{$\scriptstyle \sim$}}}}
\def\ls{\mathrel{\raise0.35ex\hbox{$\scriptstyle <$}\kern-0.6em
\lower0.40ex\hbox{{$\scriptstyle \sim$}}}}

\shortauthors{Ledlow, Owen \& Miller}
\shorttitle{Cluster of Galaxies Surrounding Cygnus A II.}

\begin{document}


\title{The Cluster of Galaxies Surrounding Cygnus A \\
  II. New Velocities and a Dynamical Model}

\author{Michael J. Ledlow\altaffilmark{1,3}} 
\affil{Gemini Observatory, Southern Operations Center, AURA, Casilla
  603, La Serena, Chile}

\author{Frazer N. Owen\altaffilmark{1}}
\affil{National Radio Astronomy Observatory\altaffilmark{2}, Socorro, NM 87801}

\author{Neal A. Miller\altaffilmark{1,4}} 
\affil{National Radio Astronomy Observatory\altaffilmark{2}, Socorro, NM 87801}

\altaffiltext{1}{Visiting astronomer, Kitt Peak National Observatory,
  National Optical Astronomy Observatories, operated by AURA, Inc.,
  under cooperative agreement with the National Science Foundation.}

\altaffiltext{2}{The National Radio Astronomy Observatory is operated
  by Associated Universities, Inc., under contract with the National
  Science Foundation.}

\altaffiltext{3}{Deceased 5 June 2004. We shall miss his cheerfulness,
unfailing good sense, and scientific industry.}

\altaffiltext{4}{Jansky Fellow of the National Radio Astronomy
Observatory, located at Johns Hopkins University, Department of Physics 
and Astronomy, 3400 N. Charles Street, Baltimore, MD 21218}

\setcounter{footnote}{4}

\begin{abstract}
  
  We have spectroscopically identified 77 new members of the Cygnus A
  cluster, bringing the total to 118 galaxies consistent with cluster
  membership.  We use these data combined with the results from X-rays
  to deduce a dynamical model for the system.  The data are consistent
  with a cluster-cluster merger viewed at a projection angle of
  30-45$^{\circ}$, 0.2-0.6 Gyr prior to core passage.  We estimate the
  richness of the combined cluster system at Abell richness
  class 2 or greater, suggesting the merger of two richness class $\sim1$
  clusters.
 
\end{abstract}

\keywords{galaxies:active --- galaxies:clusters:individual(Cygnus A) --- 
galaxies: distances and redshifts --- galaxies:individual(Cygnus A)}

\section{Introduction}

Cygnus A ($z=0.056$) is the best studied example and prototype of a
powerful FR II radio galaxy and is also one of the brightest nearby
X-ray sources.  Imaging from {\sl Einstein} showed that Cygnus A
resides within $\approx 10^{14} M_\odot$ of hot gas nearly
$1h_{75}^{-1}$ Mpc in extent \citep{arnaud84}.  Cygnus A itself has a
very high X-ray luminosity, which classically, has been associated
with a strong cooling-flow with a mass inflow rate possibly as high as
$\sim 250~M_\odot$ yr$^{-1}$ \citep{reynolds96}.  More recent
observations of cooling-core clusters with {\sl XMM} and {\sl Chandra}, 
however, have motivated both alternative explanations for the observed 
X-ray excess and imply substantially less cool gas than previously thought
for these systems \citep{bregman2003,bohringer2003}. While Cygnus A
itself has been studied extensively at all wavelengths, very little
has been known of the optical properties of the galaxy cluster or
group within which Cygnus A is found on account of its low Galactic
latitude (5$^\circ$) and attendant confusing star field.

In Paper 1 \citep{paper1} we presented 41 new velocity measurements
within a 22\arcmin{} square region surrounding Cygnus A.  Prior to
that work only 4 concordant redshifts were known \citep{spinrad}.  The
results indicated that Cygnus A was located, but not centered, in a
rich (at least Abell richness class 1), high velocity dispersion
cluster.  While clearly showing substructure suggestive of a cluster
merger, the number of velocities was insufficient to develop a
dynamical model for the system.

Using {\sl ASCA}, \citet{markevitch} mapped the gas temperature across the
cluster and showed the existence of a hot region between Cygnus A and
the secondary X-ray peak to the northwest (in the direction of the
dynamical centroid).  The temperature structure could be explained by
a fairly simple model involving the merger of two similar mass
subclusters colliding head-on and developing a shock between them.  

In this paper, we analyze new imaging and spectroscopy of the Cygnus A
cluster.  We compare the results of a detailed substructure analysis
to the cluster merger model based on the X-ray gas temperature
structure.  We discuss the new observations in $\S$2. In $\S$3.1 we
look for evidence of bimodality in the spatial and velocity
distribution of the cluster.  A full 1-D, 2-D, and 3-D substructure
analysis is presented in $\S$3.2.  In $\S$3.3 we calculate a new
estimate for the cluster richness and mass, followed by a Discussion
of results and comparison to models in $\S$4.  Conclusions are
summarized in $\S$5.

\section{Observations}

Spectroscopic observations were queue-scheduled on the WIYN 3.5m
telescope for semesters 1997B and 1998A with the Hydra-Bench
Spectrograph.  We used the 400 line grating covering the spectral
range 3880-7060 \AA\ with a dispersion of 1.55 \AA\/ pixel$^{-1}$ and
the Blue Simmons camera.  In 1997B three fiber configurations were
observed through clouds for 3x20 minute exposures on the nights of 8
and 30 September and 1 October 1997.  In 1998A five fiber configurations
were observed in clear conditions on the nights of 29 May and 28-29
June 1998.  For each fiber configuration we typically assigned 7-8
fibers to alignment stars and 20-25 sky fibers, leaving 60-70 fibers
available for galaxy candidates.

In the first 1997B WIYN queue run we targeted 166 different objects.
We used our original KPNO 0.9m 22' field centered on Cygnus A from
\citet{paper1} (Paper 1) supplemented by images from the Digitized Sky
Survey to take advantage of the larger field of view of Hydra (1
degree).  While these observations were strongly affected by clouds
and mostly unsuccessful, we were able to cull 10 new galaxy velocities
and confirmed another 4 of the lower S/N spectra from Paper 1.
Approximately 70\% of the useful spectra turned out to be stars; the
poor resolution of the Digitized Sky Survey made galaxy/star separation
very difficult in this crowded field.  Even with good quality imaging,
there is substantial confusion between blended, faint point-sources
and the appearance of an extended object, making this one of the most
challenging fields for galaxy selection.

Before the subsequent queue run in 1998A, we obtained new $R$-band
images at the KPNO 0.9m telescope on 29-30 April 1998 using the
2048x2048 T2KA CCD at a scale of 0\farcs68 pixel$^{-1}$. To complement 
the $22\arcmin \times 22\arcmin{}$ image centered on Cygnus A used 
in Paper 1, we imaged at two new positions\footnote{All positions in 
this paper are epoch J2000} (19:58:14 +40:55:43 and 20:00:06 +41:04:42) 
providing full spatial coverage over the extent of the X-ray emission. The 
two new images consisted of 4x10 min dithered exposures with a final image
quality of 1\farcs4 - 1\farcs7.  Candidate galaxies were identified using
FOCAS and inspected individually by eye. Coupled with the FOCAS
results, we used the peak-flux as an additional cut on eliminating
stars.  Priorities for spectroscopic observation were assigned primarily 
based on the optical
magnitude.  We targeted 254 galaxy candidates including many of the
measured galaxies from Paper 1 which were of low to medium S/N.
Having observed the brighter targets in 1997B (suffering more stellar
contamination) and coupled with better weather, the 1998A observations
were much more successful in finding new galaxy redshifts.  From the
1998A data we measured 88 new velocities and re-confirmed at better
signal-to-noise 15 of the more questionable objects from Paper 1. 

Velocities were measured using the IRAF cross-correlation task FXCOR
with the exception of a few emission-line objects.  Velocity errors
were calculated from the Tonry \& Davis R-value \citep{td79} as
determined by FXCOR.  For emission lines, we took the mean observed
velocity over the lines and applied a heliocentric correction.  We
assume an error of 30 km sec$^{-1}$ on emission-line velocities.

We imaged the Cygnus A field again on 19 July 1999 in $R$-band at the
KPNO 0.9m telescope, this time using the 8K x 8K Mosaic camera giving
a full one degree field of view with a scale of 0\farcs425 pixel$^{-1}$.
Details of the data reduction, photometry, and astrometry are given in
\cite{neal2003}. The final image is a sum of 5$\times$5 min exposures under
excellent conditions. The final image quality was 0\farcs85.  We used
this image in all subsequent analysis. We measured aperture magnitudes
for all new cluster galaxies using a Gunn-Oke \citep{gunnoke} metric
aperture of diameter 26.2 $h_{75}^{-1}$ kpc. Magnitudes for individual 
galaxies were calculated using a range of different procedures for handling
of foreground stars, which indicated that the subtraction of such
objects in the high stellar density field of Cygnus A was the 
primary source of error in reported magnitudes. These tests indicated
a global error of about 0.2 magnitudes for the Gunn-Oke aperture
photometry, compared to an error of 0.05 magnitudes in the
photometric solution \citep{neal2003}. Absolute magnitudes were
calculated assuming a Galactic absorption of $A_R$ = 0.81
\citep{spinrad} and a K-correction of 0.08 magnitudes, using 
$H_0 = 75$ km sec$^{-1}$ Mpc$^{-1}$ and $q_0=0.1$. For non-cluster 
galaxies (spectroscopically confirmed) and newly identified galaxies 
we used a fixed aperture of 5\arcsec{} diameter. This smaller
aperture was chosen in order to reduce the effects of foreground
stars. An identical analysis to that performed for the Gunn-Oke
photometry indicated a typical error of 0.1 magnitudes for galaxies
brighter than $m_R = 19$. All magnitudes included in tables and
figures have had foreground stars removed from their apertures.

\section{Results} 

\subsection{Velocity Distribution and Bimodality}

In Table 1 we list all new velocities measured from WIYN.  Starting
from all 139 velocities (including those from Paper 1), we calculated
the biweight estimators of location and scale \citep{beers90}.  We
then trimmed the list according to the first-pass $3\sigma$ rejection
about the central velocity and re-computed the biweight estimators on
the edited list.  This 2nd iteration produced a list of 118 cluster
members within a $3\sigma$ dispersion of the central biweight velocity
$C_{BI}=19008^{+151}_{-172}$ km sec$^{-1}$ ($z=0.0634$) and biweight
scale $S_{BI}=1489^{+123}_{-103}$ km sec$^{-1}$, corrected to the rest
frame of the cluster.  These values compare quite well to those
determined in \citet{paper1} ($C_{BI}=18873, S_{BI}=1581$ km
sec$^{-1}$), while we have increased the number of confirmed cluster
members by more than a factor of 2. Cygnus A ($cz=16811$ km sec$^{-1}$)
is therefore offset 2197 km sec$^{-1}$ from the mean cluster velocity.
For cluster members, Table 1 includes the right ascension and
declination (J2000.0), the heliocentric velocity and error estimate,
and Gunn-Oke aperture magnitudes.  We re-list velocities for 19
galaxies re-observed from Paper 1 for which we have adopted the new
velocity measured with WIYN (higher S/N). In all cases but one, the
agreement between velocities was within a few hundred km sec$^{-1}$
(typically 1-2$\sigma$).  We also list velocities for 21
background/foreground galaxies and apparent magnitudes within
a 5\arcsec{} aperture.

\begin{figure}
\plotone{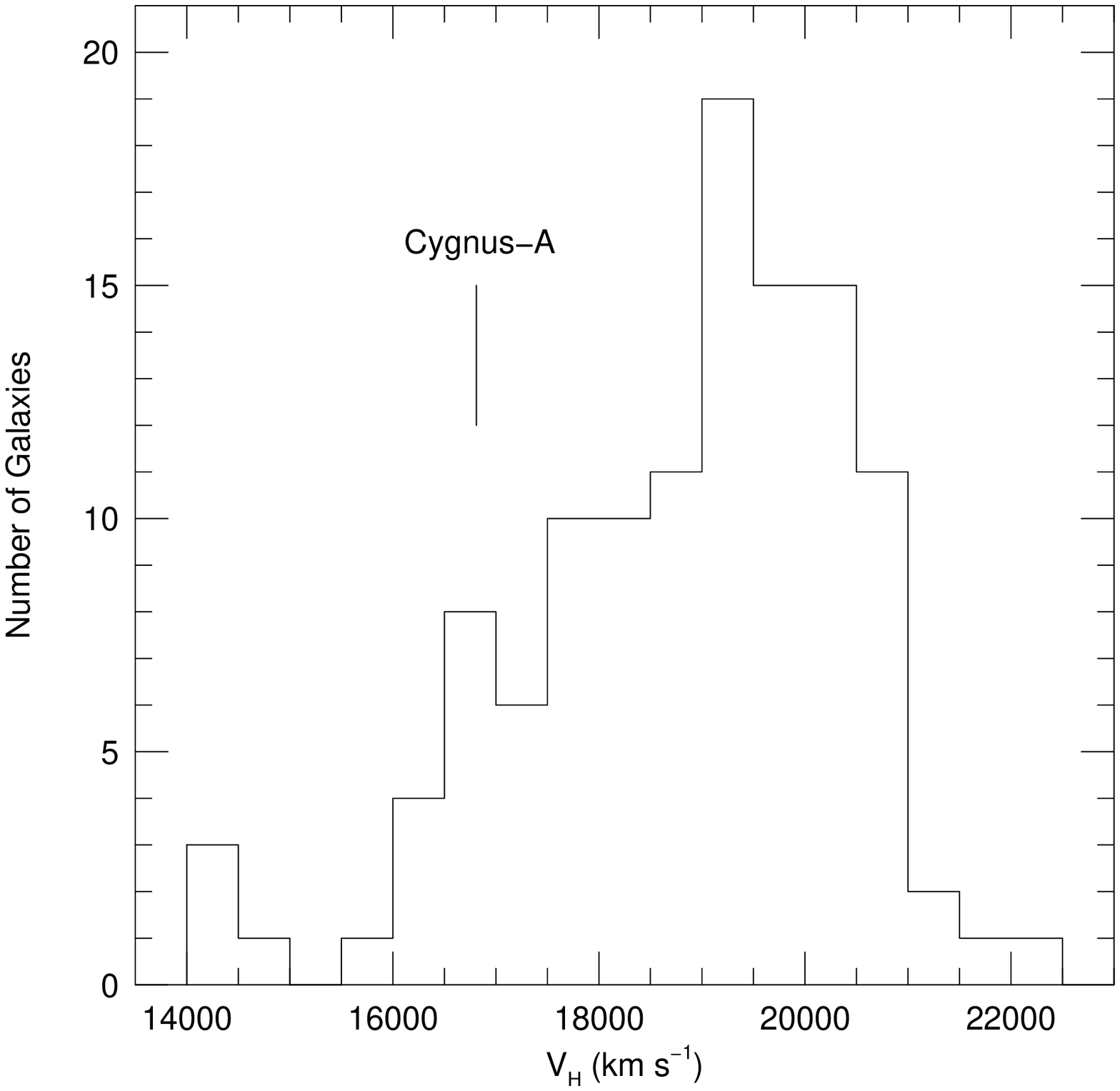}
\caption{Histogram of the radial velocities for all 118 defined
  cluster members. The binwidth is 500 km sec$^{-1}$ \label{fig1}}
\end{figure}

We plot a velocity histogram for the 118 cluster members in 
figure~\ref{fig1}.
While qualitatively similar to that in Paper 1, the bimodality
suggested by the previous work initially appears much less
significant. There is, however, clearly a heavier tail in the
distribution in the vicinity of Cygnus A.  We calculate the likelihood
of a bimodal distribution using the KMM algorithm \citep{kmm}, which
compares the goodness of fit between unimodal or multimodal Gaussian
functions and objectively partitions data into sub-populations
accordingly.  Interestingly, we find that a single Gaussian model is
rejected at a significance of $>$99\%.  The estimated velocity
locations of each group are 16648 (group 1; just 163 km sec$^{-1}$
offset from Cygnus A) and 19428 km sec$^{-1}$ (group 2).  The
partitioning assigned 23 members to group 1 and 95 to group 2. While
this result is quite significant, there was some trouble in making the
subgroup assignments with the estimates of correct allocation rates for
each group being 76\% and 96\%, respectively.  These results are for
the homoscedastic case assuming a common covariance (velocity
dispersion squared) where the estimated velocity dispersion for both groups
was 1126 km sec$^{-1}$. The results for the
heteroscedastic hypothesis (different velocity dispersions) were quite
different, suggesting both different mean velocities (18131 and 19849
km sec$^{-1}$) and mixing ratios (58 and 60 members) for the groups.
While still technically significant at the 99\% level (note that the
probability distribution is not well determined in this case so the
significance is not robust), the estimates for the correct overall
allocation rate was 78\% (73\% and 84\% for groups 1,2 respectively).
The estimated dispersions were also quite different; 1651 and 755 km
sec$^{-1}$ for groups 1 and 2, respectively. 

We also used a variant of the traditional KMM code which fits
structures based on positions as well as velocities, although it assumes
that the covariance among the three attributes is zero \citep[\ie 
there is no correlation between position and velocity;][]{a2256paper}.  
As is the case for the standard KMM, the significance of the 
multicomponent fits are not easily derived. However, a Monte Carlo 
shuffling of the velocities is used to
determine the frequency at which a randomized dataset fits as well as
the observed one.  The results, shown in figure~\ref{fig2}, were quite 
similar
to those of the heteroscedastic case given above; two groups with mean
velocities of 17648 (N=61) and 20126 km sec$^{-1}$ (N=57) and
dispersions of 1134 and 600 km sec$^{-1}$ respectively.  Note that the
velocity distribution of the higher velocity group appears much more
Gaussian-like than the more dispersed, lower-velocity group which
includes Cygnus A. In fact, statistical tests performed on this 
lower-velocity system indicate that its velocity distribution is 
non-Gaussian.

In summary, the KMM tests indicate that the Cygnus A cluster is
poorly fit by a single system. It may be represented by a pair of
systems, with one being consistent with a relaxed cluster having a
Gaussian velocity distribution and the other (which includes Cygnus A)
more difficult to describe. The velocity dispersions of these two
systems imply a virial mass ratio of over 3:1. However, based on numbers 
of assigned galaxies and the observation that the grouping containing
Cygnus A is not well described as a virialized system, these two 
systems may be of comparable richness and hence mass. A model which 
fits these characteristics will be developed in the ensuing sections.

We show the velocity-coded spatial distribution in figure~\ref{fig3}.  
Square
symbols are the galaxies with velocity less than the cluster mean while
triangles represent those at greater velocity. Shading is used to indicate
the magnitude of the difference between a galaxy's velocity and the 
cluster mean, with solid symbols for galaxies within $\pm1\sigma$ 
and open symbols for galaxies with a larger velocity differences.
For example, Cygnus A has a velocity $~1.5\sigma$ below the mean, and
hence is coded as an open-square. Note the significant grouping around 
Cygnus A, below and slightly to the left of center, which includes seven
satellite galaxies with a small dispersion in velocity.
Interestingly, the positions of the galaxies in this lower velocity bin 
(the open squares in figure~\ref{fig3}) show a linear alignment along the 
principle axis of the cluster.

\subsection{Substructure Tests} 

We ran a suite of 1D, 2D, and 3D statistical tests in order to assess
and quantify the amount and nature of substructure present in the
Cygnus A cluster.  These tests are defined and explained in detail in
\citet[][hence referred to as P96]{pinkney96}. Briefly, the 1D tests
are performed solely on the galaxy velocities, the 2D tests on galaxy
positions, and the 3D tests on both velocities and positions.

Nearly all of the 1D tests produced a significant result, with the 
main finding being that the velocity distribution is skewed.
Averaging over the 3 skewness tests, the distribution is significantly
skewed at better than the 98\% level.  The DIP-stat \citep{hartigan}
tests for consistency with a unimodal population, and returned a 99\%
significance to deviations from a single Gaussian distribution,
consistent with the KMM results discussed above.

The 2D substructure tests are useful for quantifying clumpiness,
asymmetries, and elongations in the spatial distribution of galaxies
in the cluster.  In figure~\ref{fig4} we show adaptively smoothed
contours of
the galaxy distribution overlaid onto the X-ray emission from the
{\sl ROSAT} PSPC. While similar qualitatively to its counterpart in 
Paper 1,
a much more obvious clump of galaxies is now seen near but not
centered (offset=72 $h_{75}^{-1}$ kpc) at the position of Cygnus A.
The two peaks in the spatial distribution are located at
19:59:30.6 +40:45:02 and 19:59:03.1 +40:49:36 with a separation of
458$h_{75}^{-1}$ kpc.  Similarly, the galaxy distribution traces well
the X-ray gas, although this is partly due to selection effects as our
imaging data were centered at locations optimized to span the full
extent of the intracluster medium (we did target candidates outside
this area based on the POSS). Note that the 2nd peak in the X-ray
emission (difficult to see with this stretch and in contrast to Cygnus
A), is located near 19:58:50.4 +40:52:14 at a projected separation of
$\sim$700$h_{75}^{-1}$ kpc.  From the 2D substructure tests, all but
the AST test (sensitive to non-central clumping) returned significant
results (all $>99\%$), indicating that the spatial distribution is
significantly elongated (Fourier elongation test), asymmetric
($\beta$-test), and bimodal (Lee2D).

The 3D substructure tests have the null hypothesis of constant mean
velocity and dispersion with position.  Consistent with the 1D, 2D,
and KMM results, we find evidence for significant substructure (Lee 3D,
99.4\%; $\Delta$, 99\%). The remaining 3D tests were less significant
than $\Delta$ and Lee-3D ($\alpha$, 94\%; $\alpha_{var}$, 89\%;
$\epsilon$, 93\%).  We show in figure~\ref{fig5} the standard $\Delta$ 
bubble
plot; the symbol size is proportional to the actual $\delta$ for each
galaxy as calculated by the Dressler-Shectman test, \citep{dressler}.
From comparison to figures~\ref{fig3} and~\ref{fig4}, we confirm that 
the clumping
responsible for the positive signal (local velocity and/or dispersion
are different from global values) is located exactly on the galaxy
clump associated with Cygnus-A.

\subsection{Mass and Richness} 

We calculated global mass estimates using a projected mass estimator
($M_{PME}$) and the virial theorem ($M_{VT}$), with values of
$M_{PME}=4.4\times 10^{15} M_\odot$ and $M_{VT}=3.0\times 10^{15}~
M_\odot$.  The larger $M_{PME}$ is consistent with the findings of P96
when there is a significant projected separation between two merging
clusters at an epoch prior to core crossing.  P96 found in general
that these estimators over-predict the real cluster masses, which are
additionally affected by the presence of merging subclusters.

In figure~\ref{fig6} we plot a histogram of the absolute magnitude 
distribution
of cluster galaxies.  The magnitudes ($M_{R}$) were measured within a
Gunn-Oke aperture of diameter 26.2$h_{75}^{-1}$ kpc, on which system 
$M_{R}^{*}\approx -22.0$. From inspection 
of figure~\ref{fig6} , one can see our incompleteness arises around 
$M_{R}^{*}\approx -20.5$ due to sensitivity limits and our ability to 
pick out these fainter galaxies for target selection.

In order to get a better estimate for the richness, we have taken
advantage of the sub-arcsecond image quality on our $R$-band Mosaic frame
to locate galaxies not yet identified spectroscopically. We were
fairly conservative in classifying objects as galaxies. Over the 
entire field, we identified 362 additional galaxies which we list in
Table 2. In figure~\ref{fig7} we show the spatial distribution of these
newly identified galaxies with adaptively smoothed contours as in
figure~\ref{fig4} but on a larger scale covering the full Abell radius 
mapped
by our mosaic image. Note that the primary cluster concentration from
figure~\ref{fig4} coincides with the peak offset slightly to the east
of center
in the full-field plot of figure~\ref{fig7}.  There is also
correspondence with
the two lowest surface-density peaks in figure~\ref{fig4} with the 
other two
concentrations in figure~\ref{fig7}.  We indeed see that the missed galaxy
distribution traces the same clustering pattern as the cluster
members. There is an additional chain of less-dense concentrations from the
southeast to southwest not sampled by our spectroscopy.  These results
suggest that overall, our spectroscopic sampling is quite good.

We followed the general prescription of \citet{abell} to estimate the cluster
richness. The Mosaic frame nearly exactly matches one Abell radius 
($R_A \equiv 1\farcm7 /z$, or 26\farcm8 for $z=0.0634$). Including
the galaxies from Table 1, we found 394 galaxies within an Abell 
radius of Cygnus A (19:59:28.3 +40:44:02). Of these, 133 lie within
the range $m_3$ to $m_3 + 2$ defined by Abell as the measure of
richness. In the original paper, Abell estimated contamination due
to foreground and background galaxies using nearby field counts.
Lacking a direct analog, we simply estimated field galaxies using
$N = \mathcal{N}10^{0.6m}$ where $\mathcal{N}=1.26 \times 10^{-5}$
galaxies steradian$^{-1}$ \citep[see][]{neal2003}. The corrected
count is then 106, placing the Cygnus A cluster in the middle to high
range of richness class (RC) 2 clusters. We note that this estimate for
contamination by foreground/background objects is consistent with our
spectroscopy results, where 118 of 139 galaxies were shown to be
cluster members. An even higher richness is quite possible, given our 
fairly conservative cut on galaxy identification and the likelihood that
additional galaxies are hidden by the numerous Galactic stars.
Lastly, if the Cygnus A system is an ongoing merger of two clusters of 
similar mass and richness, this would imply that they are each rich 
systems of RC=1.

\section{Discussion} 

\subsection{Optical Results: A Dynamical Model} 

The picture gleaned from the substructure analysis suggests that
Cygnus A is the dominant galaxy in a cluster which is currently
merging with another subcluster.  The fact that we get a consistent
picture from each of the KMM, 1D, 2D, and 3D tests allows us to
constrain somewhat the expected merger properties.  

\begin{figure}
\plotone{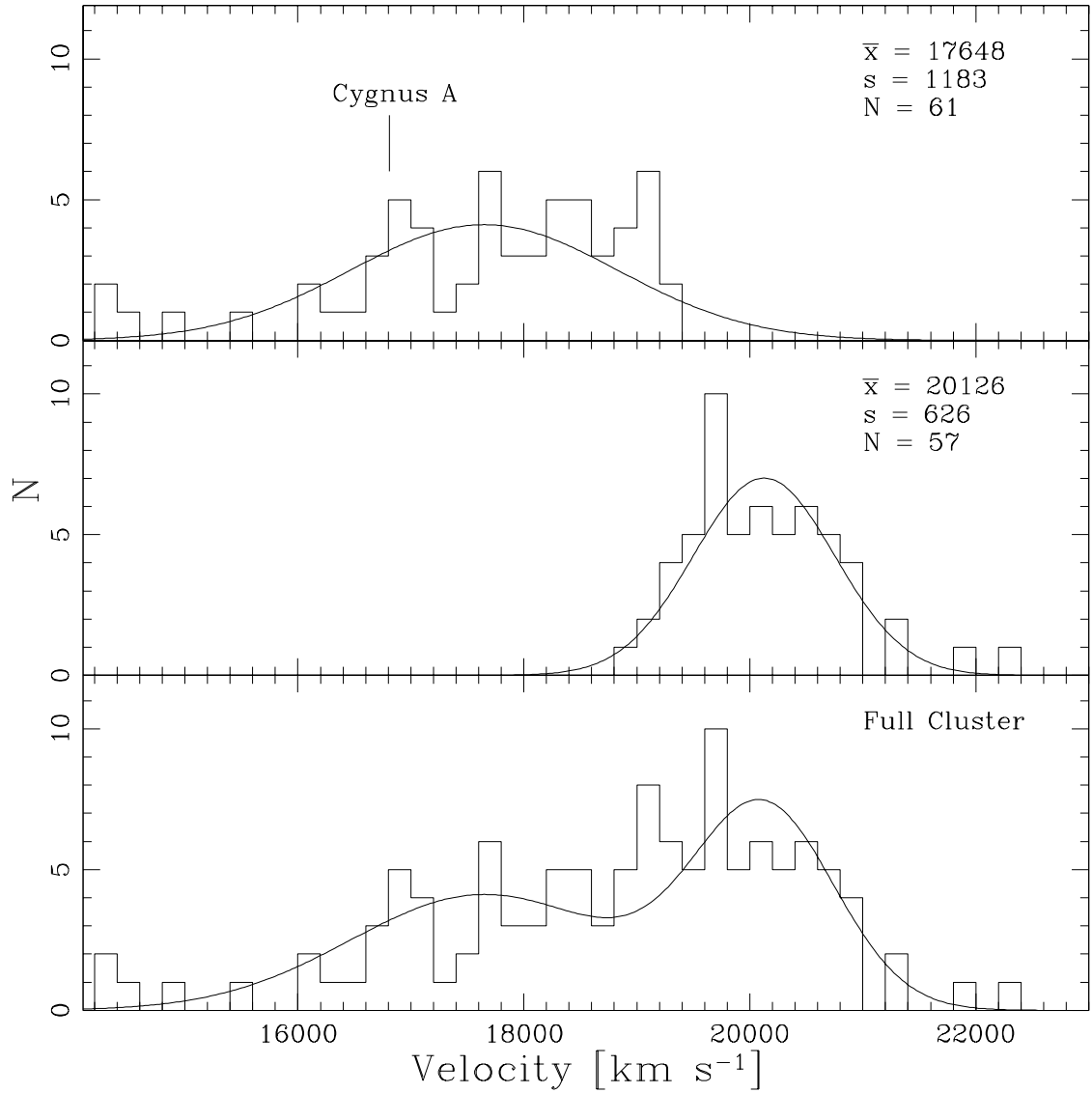}
\caption{Histograms and Gaussians representing the fitted
  location and dispersions of the two subclusters as separated by the
  KMM algorithm.  The lower panel shows the composite velocity
  distribution and fits.  A velocity binwidth of 200 km sec$^{-1}$ was
  used. \label{fig2}}
\end{figure}

\begin{figure}
\epsscale{.90}
\plotone{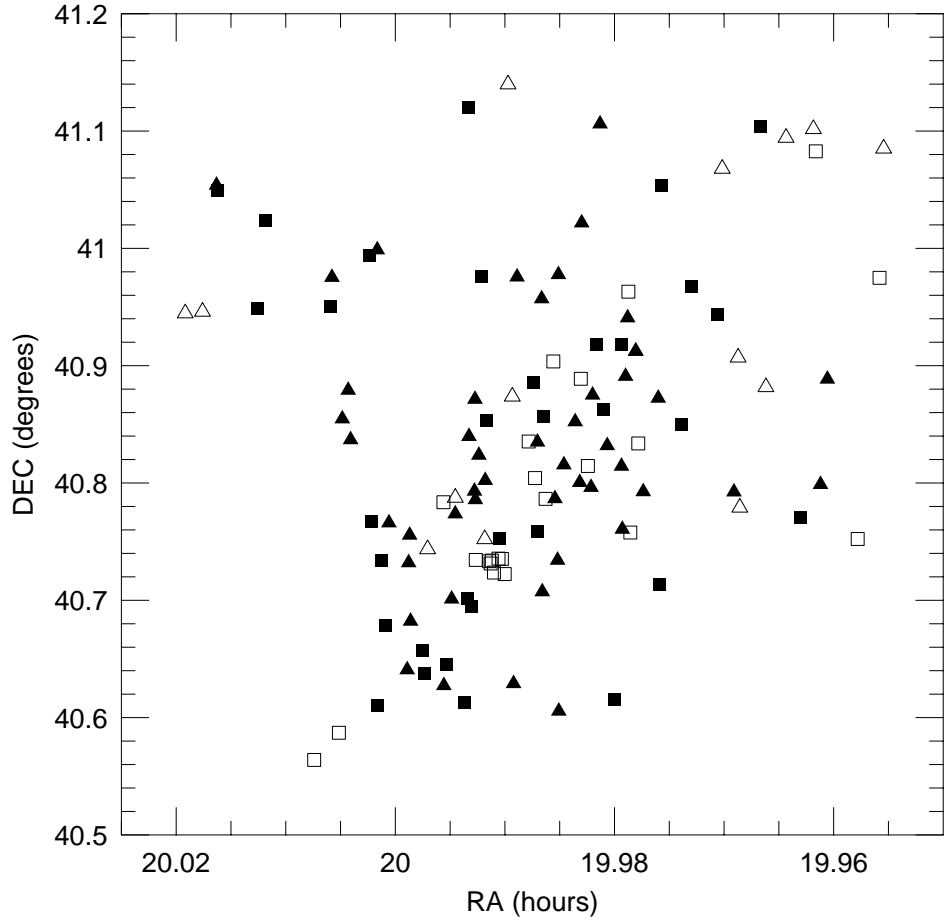}
\caption{Velocity-coded spatial distribution of cluster members.
  Solid symbols are shown for galaxies with velocities within
  1$\sigma$ of the mean (biweight location), open symbols for
  velocities offset more than 1$\sigma$. Squares indicate a velocity
  less than the mean, triangles greater than the mean.  The
  concentration of open-squares below and to the left of center is the
  Cygnus-A grouping. \label{fig3}}
\end{figure}

\begin{figure}
\plotone{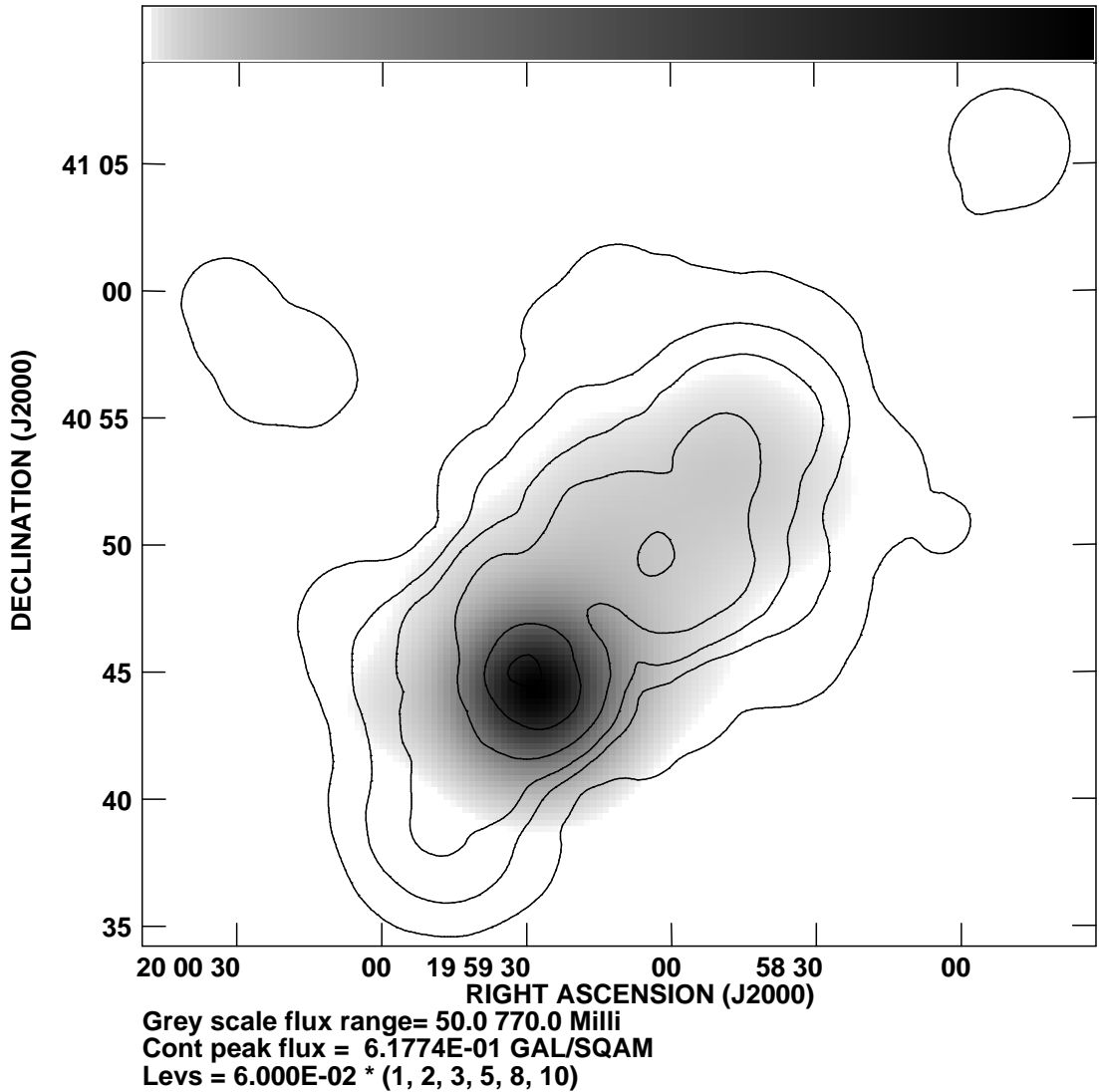}
\caption{Adaptively smoothed galaxy distribution of
  spectroscopically confirmed cluster members given in Table 1
  (contours), superposed on the ROSAT PSPC X-ray image (greyscale).
  The galaxy density contours are 1,2,3,5,8, and 10 $\times$ 0.06
  galaxies arcmin$^{-2}$. Cygnus A is located at the X-ray peak of the
  greyscale image. There is a secondary X-ray peak to the northwest
  which is difficult to see with this stretch. Its location is noted
  in the text. \label{fig4}}
\end{figure}

\begin{figure}
\plotone{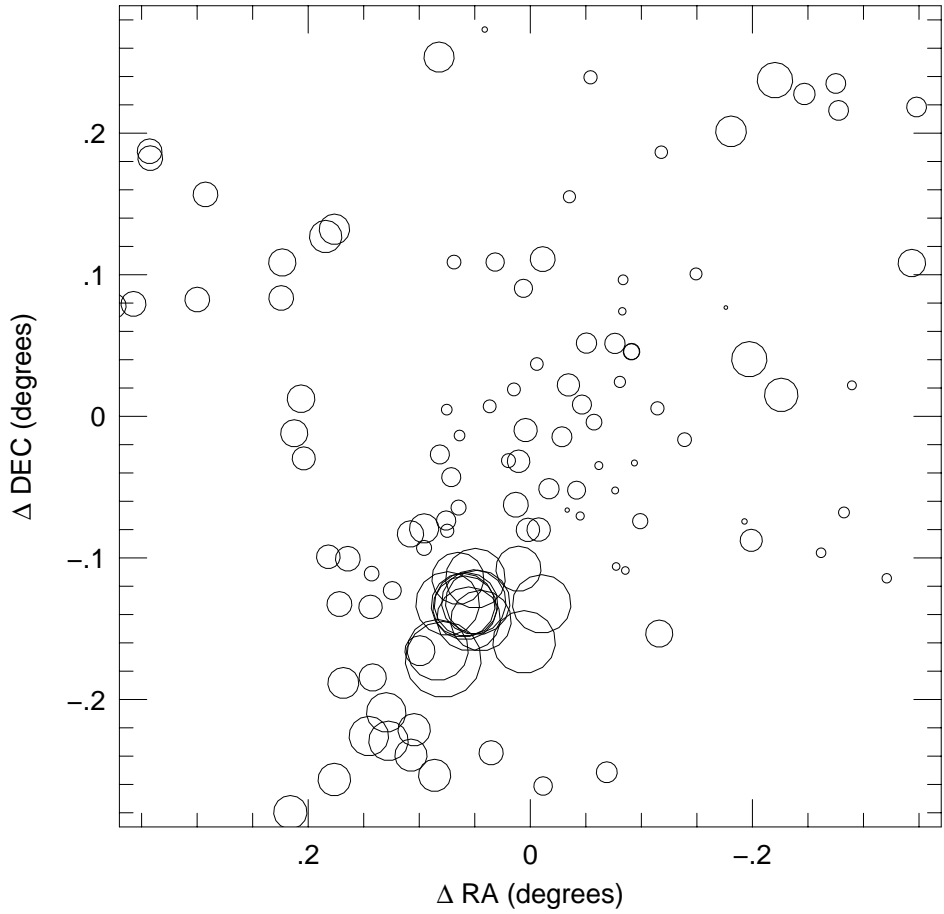}
\caption{Dressler-Shectman bubble-plot showing the results of the
  $\Delta$-test for 3D substructure.  The size of the circles are
  proportional to the $\delta$-value calculated for each galaxy. The
  strong positive signal of subclustering is centered on Cygnus A.
 \label{fig5}}
\end{figure}

\begin{figure}
\plotone{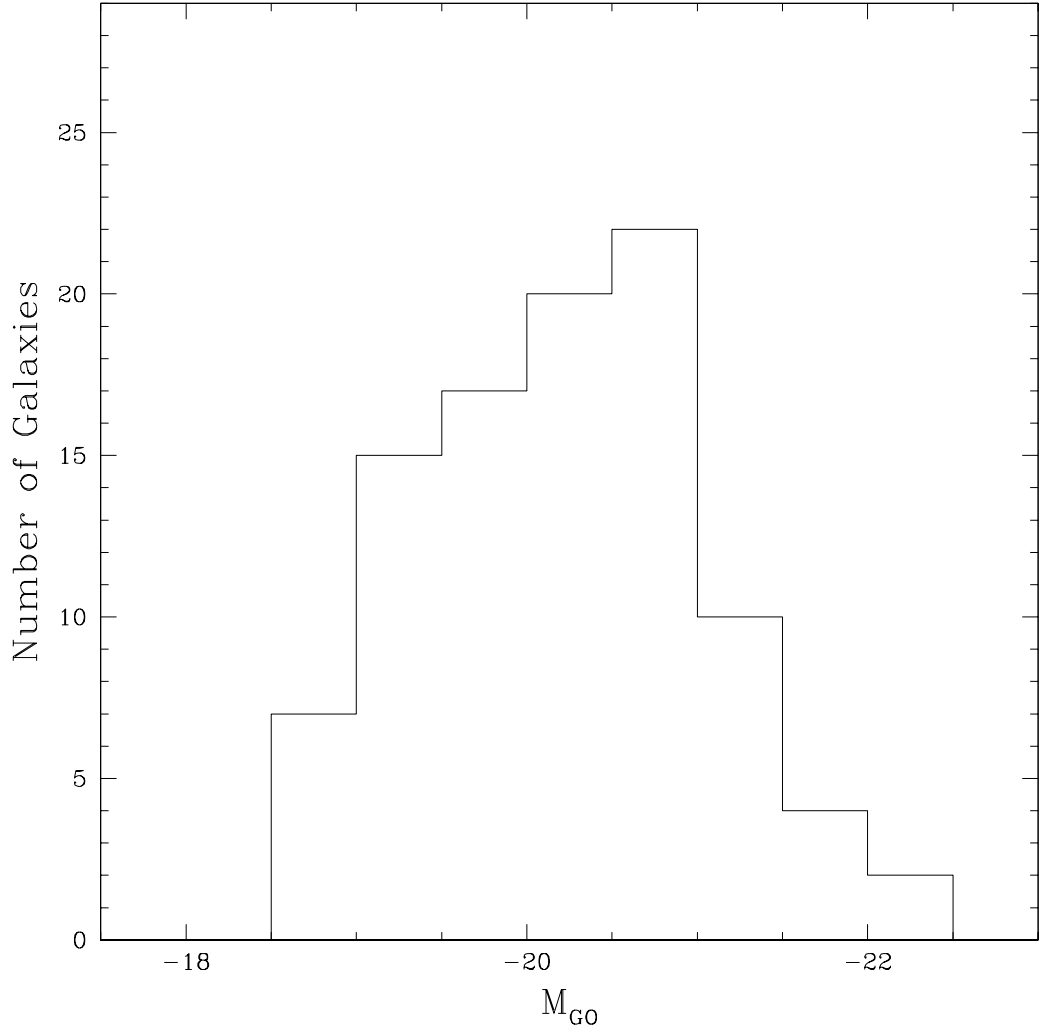}
\caption{Histogram of absolute magnitudes measured within a
  Gunn-Oke aperture of metric diameter 26.2$h_{75}^{-1}$ kpc for the
  118 confirmed cluster members \label{fig6}}
\end{figure}

\begin{figure}
\plotone{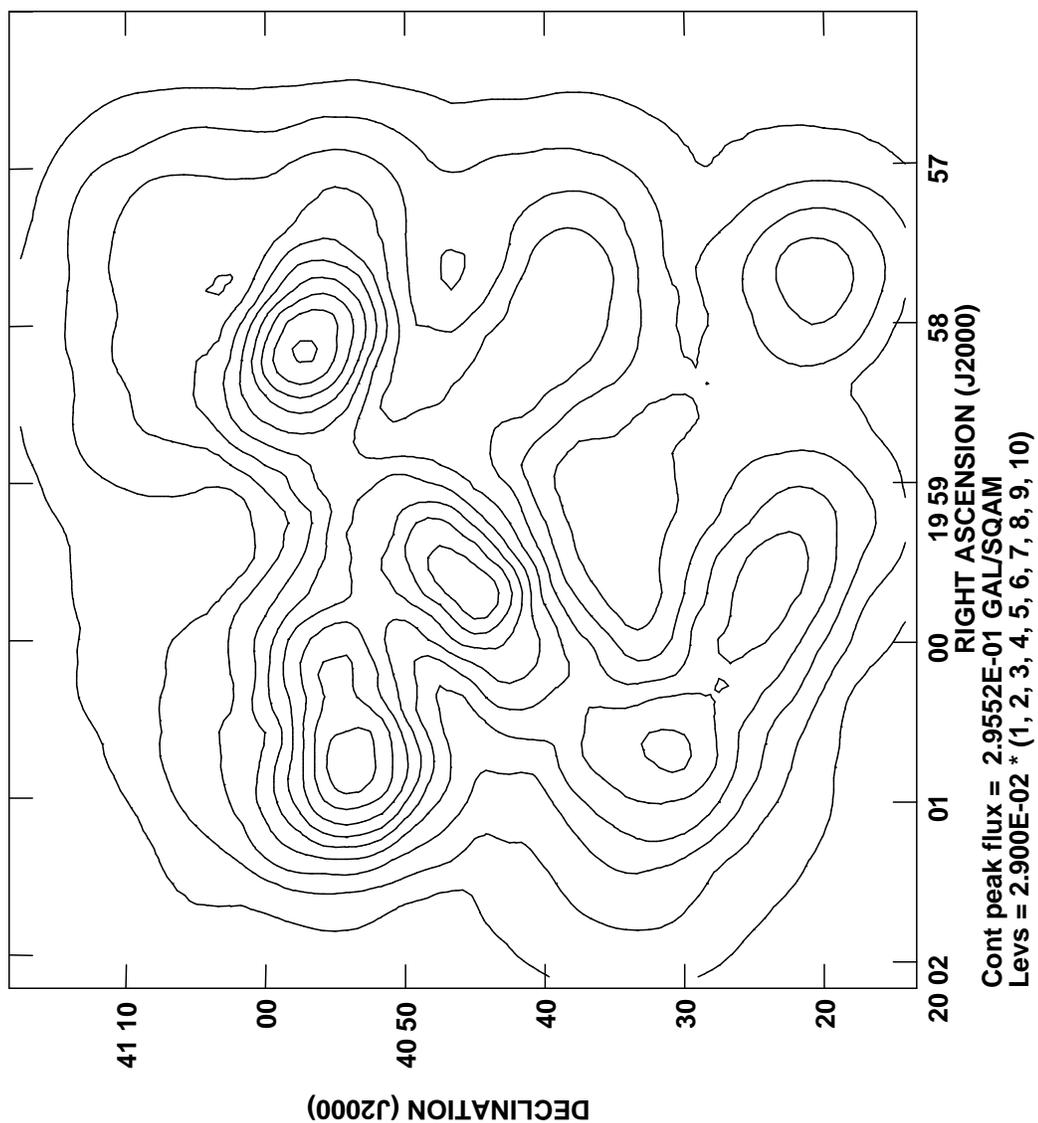}
\caption{Adaptively smoothed contours of the spatial distribution
  of 362 galaxies identified from the $R$-band Mosaic which were not
  targeted by spectroscopy. While many of these certainly are
  background/foreground to the cluster, the surface density in fact
  matches quite well with our spectroscopic sample (Figure 4),
  suggesting that our spectroscopic sampling is not strongly biased.
 \label{fig7}}
\end{figure}

Depending on which variant of the KMM partitioning scheme is used we
find that the velocity separation of the two subclusters is between
1600 and 2600 km sec$^{-1}$ (corrected to the rest-frame of the
cluster).  We find different mixing ratios for the two cases
(homoscedastic or heteroscedastic assumption); with equal velocity
dispersions the ratio of galaxy members is of order 4 (Cygnus-A being
the poorer group) whereas they are nearly equal when the dispersions
are fit independently. The projected separation of the subclusters is
$\sim$700$h_{75}^{-1}$ kpc from the X-rays or $\sim$460 $h_{75}^{-1}$
kpc based on the galaxy surface density peak.  We compare the
properties of the two subclusters using
figures~\ref{fig2}-~\ref{fig4}.  The KMM fits
suggest that the Cygnus A subcluster is fairly dispersed in velocity
whereas the higher velocity peak is more Gaussian.  As seen in
projection, spatially the two subclusters appear to be well-mixed,
although the immediate clumping centered on Cygnus-A is quite
significant with a small velocity-spread, well offset from the global
mean.  And as previously noted, the KMM group which includes Cygnus A
also shows an interesting linear alignment between the two
subclusters, possibly coincident with the axis of the merger.  

Now, we utilize the N-body results and statistics from the
substructure tests in P96 to infer a best model for the merger state
of this system. P96 note that the 1D, 2D, and 3D tests each have
their own strengths and weaknesses so a collective interpretation
of their results can reveal an estimate of the merger geometry.
The 1D substructure tests are obviously most 
sensitive when viewed along the merger axis (in our discussion,
this is called a projection angle of 0$^\circ$), and fail to produce 
significant results for mergers viewed at projection angles greater 
than $\approx60^\circ$. The 2D tests require a larger projection 
angle to separate the components, although the size of this angle 
required to produce significant substructure tests is dependent on 
the epoch of the merger. Projection angles smaller than 
$\approx30^\circ$ only produce significant results well before or 
after core passage. The 3D tests are most sensitive to projection
angles $\ls60^\circ$, similar to the 1D tests. Thus, producing
significant substructure results for all three sets of tests
requires an intermediate projection angle of $\sim$30-45$^{\circ}$.
Furthermore, the simulations indicate that larger mass ratios decrease
the significance of many tests, further implying a mass ratio of less
than 3:1 for the two subclusters. Specifically, the degree of significance 
of the 2D and 3D substructure tests we observe is only present in the 
simulations at times $\sim$0.2-0.6 Gyr prior to core passage, or else
significantly after core passage ($>$2 Gyr). The high velocity dispersion 
and 3D tests strongly favor the pre core-passage epoch.  The 1D results also
support a time quite near to core crossing, but do not discriminate
between the pre or post-crossing event.  The subcluster morphologies
are interesting in this context; the Cygnus A subcluster appears
dispersed (figure~\ref{fig2}) as if the core of the subcluster were lagging
behind its member galaxies in the merger. Also note the compact core
around Cygnus A, which is responsible for the significant
$\Delta$-test results (figure~\ref{fig5}).  Post core-passage, 
it would seem
unlikely that such a core would remain so distinct.  The other
subcluster appears more Gaussian in velocity space, is quite well
mixed spatially with the Cygnus A subcluster, and lacks a well-defined
core.  It is not clear which of the subclusters is the {\it primary}
in the merger, although the data do not suggest a large mass ratio so
the distinction may be irrelevant.

\subsection{The Picture from X-rays} 
 
More details about the likely merger in this system can be found from
properties of the X-ray gas.  \citet{markevitch}(MSV99) studied X-ray
gas temperature maps from {\sl ASCA} for several nearby merging galaxy
clusters, including Cygnus A.  MSV99 mapped the X-ray temperature
across the cluster from an annulus including the Cygnus A group
(Cygnus A was excised from the data before fitting the temperature) in
the direction of the 2nd spatial peak (see figure~\ref{fig4}).  The results
indicate a region between the two clusters with T$\approx 8-9$ keV
whereas both subclusters are similar at $T=4-5$ keV. The hot region is
consistent with a shocked region resulting from a head-on collision of
similar mass clusters. Under this fairly simple geometry and under the
approximation of a 1-dimensional shock, MSV99 estimate a subcluster
collision velocity of $2200^{+700}_{-500}$ km sec$^{-1}$.  This value is
close to the expected free-fall velocity that two similar mass, $T\sim
4-5$ keV, clusters would achieve by the time that they reached the
observed separation.  While this simple 1D shock model ignores the
physics of the gas in the interaction, MSV99 confirm that the velocity
estimate is accurate to about 20\% as compared to when hydrodynamical
effects are included in the model.  Quite interestingly, this
predicted velocity matches extremely well the velocity difference of
the two subclusters inferred from the optical data (1600-2600 km
sec$^{-1}$).

More recently, \citet{smith2002} studied the ICM surrounding Cygnus A
with {\sl Chandra}.  While the more diffuse X-ray emission extending
to the northwest and the 2nd cluster is partially detected on two of
the other chips, the sensitivity is insufficient to characterize its
properties. For the southwestern half of the X-ray emission in
figure~\ref{fig2} , however, Smith et al. confirm the {\sl ASCA} 
temperature results
and compute both the gas mass and total integrated cluster mass.
Within a 500 kpc radius they derive a gas mass of $1.1\times 10^{13}
M_\odot$ and total mass between $2.0-2.8\times 10^{14} M_\odot$
(depending on the central temperature profile). We estimate a factor
of 4-5 higher total mass, based on the projected mass from the optical
data. However, it is hard to directly compare the two estimates since
Smith et al assumed a lower Hubble constant
(\ie $H_0=50$ km s$^{-1}$Mpc$^{-1}$), and  
the {\sl Chandra} measurement includes about 1/3 of the full
area of the galaxy distribution and extended X-ray emission. 
\cite{smith2002} also show a complex temperature distribution for
the cluster consistent with a unsettled dynamical situation in the
hot gas. 

We looked for counterparts to the X-ray point sources detected in the
{\sl Chandra} image of \citet{smith2002}.  With the exception of Cygnus
A itself, we find no identifications with our extended list of
galaxies in Tables 1 and 2, which confirms their conclusion that these
sources are most likely stars or background objects (\eg AGNs).

\subsection{The Cooling-Core and a Cluster Merger?} 

One other observation to add to this picture is the presence of the
cooling-core, \ie a core with a cooling time much less than a Hubble
time, in Cygnus A.  While the interpretation of so-called
cooling-flow clusters appears to be undergoing revision
\citep{bregman2003}, some aspects of the more classical picture may
survive. Cygnus A is a central-dominant galaxy with a central X-ray
peak, a negative or flat temperature gradient and positive metallicity
gradient toward the core, and presumably sits at the bottom of the
local gravitational potential well \citep{smith2002}. Thus one would
expect significant cooling to occur in its core. However, the X-ray
spectrum of the core shows a strong power-law continuum to very high
energies as expected for a powerful AGN/Quasar \citep{young}.
According to current models, it is expected that the powerful radio
source may have a significant influence on the cooling properties of
the central gas, with the result that the mass deposition rates
inferred by the central X-ray excess alone may be over-estimations by
more than an order of magnitude
 \citep{fabian2003,bohringer2003,bregman2003}.

Current observations of cooling flows show that high metal abundances
are common and thus require large enrichment times to create them
\citep{bohringer2003}.  The implication is that cooling cores probably
form very early in clusters, are very persistent phenomenon, and
likely survive cluster mergers as well as the energy input over the
lifetime of the central AGN's.  So most mergers must not destroy
cooling cores. Consistent with this hypothesis are the observed high
frequency of cluster substructure and the frequency of cooling cores
seen in large samples of clusters \citep[e.g.][]{jf99}. Disturbances 
may occur as a result of a merger,
but may only take the form of a {\it sloshing} of the central gas.
\citet{churazov2003} observed the signs of such an event in the
Perseus cluster.  These authors also argue that the cold, dense,
low-entropy core expected to be found in Perseus (and similarly in
Cygnus-A) is probably well protected against the penetration of gas
from infalling groups or clusters in a merger event.

In an alternative hypothesis, \citet{burns2003} suggest that cores of
cool gas first form in subclusters which then merge to create rich
clusters with cool, central X-ray excesses.  In this model, cool cores
actually form in hierarchical formation along with the clusters
themselves. This model is attractive from the point of view of
providing sufficient time to produce the observed central
metallicities and is consistent with the presence of substantial
substructure in these clusters.

Following the more classical picture of cluster cooling flows,
\citet{gomez} studied the consequences of head-on 4:1 and 16:1
mass-ratio mergers on existing cooling-flows.  Their results indicated
that the key parameters to whether the cooling-flow survives such an
event were the gas dynamics in the interaction of secondary and
primary cluster gas.  Those mergers with higher secondary gas
densities were found to be the most disruptive.  However, even in the
cases of strong disruption of the cooling flow, their simulations
indicate that the initial increase in the depth of the gravitational
potential at core-crossing is insufficient to disrupt the flow.  The
demise of the cooling flow in these simulations occurred after a
time-delay of 1-2 Gyr following core-crossing.  Applied to the Cygnus
A cluster, the properties inferred from both optical and X-ray do not
suggest a dense cluster core for the higher velocity subcluster.  So
even for a high mass density in the secondary core (which is not
supported by the
data), the effects on the cooling flow are expected to be minor. Thus
the presence of the cooling core in Cygnus A does not argue strongly
either way for a pre or post core-crossing epoch.  In any case, one
would not expect significant disruption, even with the classical
cooling flow model.

\subsection{A Connection to the Radio Source and AGN?} 

The existence of the most powerful radio galaxy in the local universe
in the midst of a massive, complex, cluster-cluster merger close to
core passage, raises the question of whether there is a connection
between the two phenomena. Radio sources are believed to be
transients. A soft lower limit to the age of $\sim 10^{6.8-7}$ can be
set from synchrotron spectral aging analysis
\citep{carilli96,carilli91}. A more complete physical argument yields
an age of  $\sim 4 \times 10^7$ years \citep{b96}. Thus the current
event we are observing seems to have an age shorter than the merger
timescale.

However, the dynamics and the morphology of the optical cluster
we report in this paper suggest a complex interaction is underway.
The X-ray total intensity and temperature imaging add to the already
complicated picture. As can be seen in figure~\ref{fig2}, Cygnus A is
not at the
center of the velocity field of the subcluster to which it is assigned
membership. Furthermore, as can be seen in figure~\ref{fig3}, Cygnus A
is not
close to spatial center of its subcluster either. However, from 
figure~\ref{fig4}, one can see that it is at the peak of the galaxy
counts as well as
the the X-ray peak. One does not see the symmetry one might expect
based on what we see in other clusters with dense cores.

Thus while the timescales of the merger event do not clearly match the
radio event, the large scale environment around Cygnus A does seem
unusually disturbed. This situation allows for a variety of physical
phenomena which could disturb the Cygnus A galaxy and produce the
radio event. Given the large velocities for the individual galaxies
and the dense environment around Cygnus A, a major merger of galaxies
seems unlikely \citep[e.g.][]{g03}. However, a galaxy in a stable orbit
around the Cygnus A might be perturbed by a close passage
and fall into the central galaxy. Recently, from adaptive optics
imaging, \citet{canalizo} have found a secondary point source located
just 0\farcs4 (400 pc in projection)
southwest of the radio core.  The spectral energy distribution of the
object is most consistent with the core of a gas-stripped galaxy,
likely merging with the giant elliptical host.  In comparison to
models, these authors speculate that the core of such a galaxy
(100-1000 times less massive than the host) might survive a few
pericenter passes or a radial encounter, the time-frame for such
events being similar (a few $\times 10^7$ years) to the estimated age
of the radio source.

 Also the cluster-cluster merger will produce
a time- variable tidal field on galaxies or dense clouds near Cygnus A
\citep{g03}. This could well disrupt a stable environment near Cygnus A 
and cause  material to fall into the core. Infalling HI gas is seen
against the nucleus \citep{cb95} and recently near-IR spectroscopy has 
revealed a giant molecular cloud also falling into the core of Cygnus A. 
\citep{b04}. Thus indirectly the complex cluster environment could be
responsible for the extreme AGN we see in Cygnus A. 

\subsection{A Coherent Picture} 

The picture which emerges from both the optical and X-rays is a
cluster-cluster merger seen in projection at 30-45$^{\circ}$ fairly
near to core-crossing, The 2D and 3D substructure tests argue for a
pre-crossing time-frame, with the level and type of substructure
consistent with epochs between 0.2-0.6 Gyr prior to core passage.

The X-ray results from the literature, as discussed in section 4.3, 
indicate the presence of a bow-shock located between
the two subclusters with a temperature enhancement a factor of two
higher than the subcluster ICMs.  A shock with these conditions is
expected to form by gas moving at supersonic speeds as two $\approx$
equal mass clusters come together and reach a separation of $\sim$1
Mpc.  Such a shock forms when the leading edge of the infalling
cluster is still several hundreds of kpc from the primary cluster
core.  As the cores come closer together in time, the shock structure
would actually become more difficult to discern because of the steep
density profile near the core (an issue of contrast).  Additionally,
the shock will decelerate as the cores merge because of the increase
in the ambient gas density.  Hence the X-ray results would also seem
to argue for a pre core-crossing epoch.

\section{Conclusions} 

Cygnus A resides near but offset from the center of a RC $\geq$2
cluster which appears to be merging with another similar richness
cluster.  The implied mass ratio of the two subclusters is of order
2:1 at most, with the Cygnus-A clump being somewhat more massive.

We have spectroscopically identified 74 new members of the Cygnus A
cluster, bringing the total to 118 galaxies consistent with cluster
membership based on 3$\sigma$ clipping of the velocity distribution.
The cluster has a biweight mean velocity $C_{BI}= 19008^{+151}_{-172}$
 km sec$^{-1}$ and biweight scale $S_{BI}=1489^{+123}_{-103}$
km sec$^{-1}$ (corrected to the cluster rest frame).

Results from the optical 1D, 2D, and 3D substructure analysis indicate
significant bimodality consistent with the merger of two subclusters.
The level of substructure seen and the specific tests which detect it
are consistent with a pre core-passage epoch, 200-600 Myr from
core-crossing.  Such a model is independently supported from the X-ray
temperature variations and the presence of shock heated gas between
the subclusters. The large velocity dispersion of the KMM substructure
associated with Cygnus A, the linear spatial distribution of galaxies
along the likely merger axis, and the existence of shocked gas between
the two cores suggests that we are viewing this system shortly before
core passage.

\acknowledgements

We thank Dr. Elizabeth Rizza for her help with the 1998 imaging data.
This research was supported by the Gemini Observatory, which is
operated by the Association of Universities for Research in Astronomy,
Inc., on behalf of the international Gemini partnership of Argentina,
Australia, Brazil, Canada, Chile, the United Kingdom and the United
States of America.

\clearpage
\clearpage
\clearpage
\clearpage
\clearpage
\clearpage
\clearpage
\clearpage
\begin{deluxetable}{r r r r r}
\tablecolumns{5}
\tablecaption{Galaxy Velocities and Absolute Magnitudes\label{tbl-vels}}
\tablewidth{0pt}
\tablehead{
\colhead{RA} & \colhead{Dec} & \colhead{$v_H$} &
\colhead{$\Delta v$} & \colhead{$M_R$} \\
\colhead{(J2000)} & \colhead{(J2000)} & 
\colhead{[km s$^{-1}$]} & \colhead{[km s$^{-1}$]} & \colhead{(GO)}
}
\startdata
19:57:19.57 & 41:05:06.5 & 20986 &  51 & -20.6 \\
19:57:20.93 & 40:58:29.6 & 17148 &  42 & -21.4 \\
19:57:28.10 & 40:45:08.2 & 14330 &  88 & -19.5 \\
19:57:38.06 & 40:53:18.8 & 19786 &  36 & -21.7 \\
19:57:40.32 & 40:47:55.6 & 20146 & 132 & -19.2 \\
19:57:41.86 & 41:04:57.8 & 14480 & 112 & -19.3 \\
19:57:42.66 & 41:06:06.3 & 20806 &  49 & -21.3 \\
19:57:46.87 & 40:46:13.1 & 18887 &  92 & -19.6 \\
19:57:51.63 & 41:05:39.6 & 20746 &  59 & -20.9 \\
19:57:58.25 & 40:52:54.4 & 20596 &  47 & -21.2 \\
19:58:00.04 & 41:06:14.7 & 17598 &  64 & -19.3 \\
19:58:06.83 & 40:46:44.6 & 20746 &  30 & -20.8 \\
19:58:07.38 & 40:54:25.2 & 20746 &  88 & -20.2 \\
19:58:08.75 & 40:47:32.7 & 19936 &  55 & -19.8 \\
19:58:12.60 & 41:04:04.5 & 22245 &  80 & -18.9 \\
19:58:14.08 & 40:56:36.6 & 18797 &  90 & -21.2 \\
19:58:22.60 & 40:58:02.1 & 18197 &  80 & -20.8 \\
19:58:25.88 & 40:51:00.6 & 17748 &  47 & -20.8 \\
19:58:32.53 & 41:03:11.6 & 18257 &  44 & -21.7 \\
19:58:38.55 & 40:47:33.4 & 19786 &  48 & -20.4 \\
19:58:40.98 & 40:54:44.9 & 19427 &  65 & -20.5 \\
19:58:43.44 & 40:57:47.1 & 14330 &  77 & -19.3 \\
19:58:43.67 & 40:56:27.0 & 20236 &  86 & -19.5 \\
19:58:44.36 & 40:53:27.5 & 20446 &  60 & -20.0 \\
19:58:45.78 & 40:55:05.4 & 18557 &  82 & -20.5 \\
19:58:50.38 & 40:49:54.8 & 19606 &  37 & -20.5 \\
19:58:51.75 & 40:51:45.2 & 17658\tablenotemark{a} &  30 & -18.6 \\
19:58:52.75 & 41:06:22.0 & 20476 &  72 & -20.4 \\
19:58:53.90 & 40:55:06.2 & 19007 &  70 & -20.9 \\
19:58:55.20 & 40:52:29.9 & 20146 &  57 & -19.6 \\
19:58:55.70 & 40:47:46.6 & 20056 &  54 & -20.8 \\
19:58:58.78 & 41:01:18.3 & 19247 &  62 & -20.9 \\
19:58:59.41 & 40:48:01.7 & 20476 &  66 & -20.0 \\
19:59:04.63 & 40:48:55.9 & 20566 &  58 & -20.1 \\
19:59:06.42 & 40:58:39.7 & 19996 &  92 & -19.6 \\
19:59:07.57 & 40:47:12.1 & 19307 &  79 & -19.1 \\
19:59:10.64 & 40:47:11.3 & 17148 &  73 & -19.0 \\
19:59:11.72 & 40:42:26.3 & 19157 &  79 & -19.2 \\
19:59:11.92 & 40:57:25.4 & 19457 &  84 & -21.2 \\
19:59:13.34 & 40:45:32.4 & 17748 &  75 & -19.8 \\
19:59:14.64 & 40:53:08.4 & 18018 &  96 & -20.1 \\
19:59:20.00 & 40:58:32.3 & 19157 &  56 & -21.1 \\
19:59:21.16 & 40:37:44.5 & 20446 &  83 & -19.1 \\
19:59:21.57 & 40:52:25.4 & 21825 & 111 & -18.6 \\
19:59:22.98 & 41:08:23.5 & 20626 &  69 & -21.3 \\
19:59:24.07 & 40:43:20.6 & 16459 &  82 & -19.9 \\
19:59:24.99 & 40:44:06.9 & 16908 &  62 & -20.0 \\
19:59:25.64 & 40:45:08.3 & 18377 &  73 & -18.7 \\
19:59:27.60 & 40:43:25.1 & 16608 &  64 & -20.4 \\
19:59:28.68 & 40:43:53.0 & 16878 &  78 & -20.2 \\
19:59:29.17 & 40:43:59.9 & 17388 & 134 & -19.6 \\
19:59:30.18 & 40:51:11.2 & 18767 &  92 & -19.5 \\
19:59:30.46 & 40:48:08.0 & 19786 &  72 & -19.8 \\
19:59:30.64 & 40:45:07.9 & 21345 &  86 & -19.2 \\
19:59:31.76 & 40:58:32.2 & 18947 &  56 & -19.6 \\
19:59:33.58 & 40:44:03.9 & 16968 & 159 & -19.1 \\
19:59:34.84 & 40:41:41.8 & 17928 & 116 & -19.6 \\
19:59:36.04 & 41:07:13.3 & 18227 &  65 & -20.8 \\
19:59:40.31 & 40:46:25.4 & 20206 &  61 & -19.9 \\
19:59:43.14 & 40:38:43.0 & 17778 &  56 & -20.0 \\
19:59:44.06 & 40:37:38.4 & 19277\tablenotemark{a} &  30 & -18.7 \\
19:59:50.48 & 40:38:14.8 & 18107 & 107 & -20.0 \\
19:59:51.12 & 40:39:27.2 & 18677 & 133 & -19.5 \\
20:00:04.51 & 40:44:02.6 & 19007 &  90 & -20.7 \\
20:00:05.93 & 40:59:55.6 & 20056 &  69 & -20.6 \\
20:00:08.44 & 40:59:37.2 & 17628 & 106 & -19.4 \\
20:00:15.48 & 40:52:44.7 & 19666 & 156 & -19.0 \\
20:00:18.54 & 40:35:13.5 & 17418 & 132 & -19.1 \\
20:00:20.83 & 40:58:31.1 & 19936 &  65 & -21.0 \\
20:00:21.14 & 40:57:00.9 & 18497 &  69 & -20.3 \\
20:00:26.67 & 40:33:50.2 & 16638 & 150 & -18.6 \\
20:00:42.78 & 41:01:24.2 & 18557 &  85 & -18.9 \\
20:00:45.13 & 40:56:56.9 & 17688\tablenotemark{a} &  30 & -19.9 \\
20:00:58.56 & 41:02:56.4 & 18947 &  96 & -20.0 \\
20:00:58.74 & 41:03:14.0 & 19487 &  63 & -20.6 \\
20:01:03.35 & 40:56:45.8 & 20896 &  34 & -22.1 \\
20:01:09.02 & 40:56:40.9 & 20836 &  68 & -21.3 \\
\cutinhead{Reobserved Galaxies from Paper 1}
19:58:33.11 & 40:42:47.5 & 17898 &  71 & -20.0 \\
19:58:33.63 & 40:52:20.3 & 19457 &  48 & -20.6 \\
19:58:40.20 & 40:50:01.4 & 16069 &  45 & -20.4 \\
19:58:41.08 & 40:54:44.0 & 19337 &  65 & -20.5 \\
19:58:45.43 & 40:45:38.2 & 19037 &  37 & -22.3 \\
19:58:45.71 & 40:48:51.5 & 20086 &  49 & -21.1 \\
19:58:56.74 & 40:48:52.2 & 16159 &  74 & -20.6 \\
19:58:59.07 & 40:53:19.8 & 16818 &  87 & -20.4 \\
19:59:00.92 & 40:51:08.0 & 19337 &  52 & -21.9 \\
19:59:08.09 & 40:54:13.0 & 16668 &  51 & -20.7 \\
19:59:13.30 & 40:50:05.7 & 19367 &  55 & -19.0 \\
19:59:14.10 & 40:48:15.2 & 16219 &  53 & -19.3 \\
19:59:16.19 & 40:50:07.3 & 15529 &  54 & -20.9 \\
19:59:32.55 & 40:49:25.3 & 19636 &  59 & -20.2 \\
19:59:33.83 & 40:52:17.0 & 19996 &  43 & -20.2 \\
19:59:55.62 & 40:43:55.5 & 20266 &  98 & -20.3 \\
20:00:02.10 & 40:45:57.8 & 19067 &  36 & -20.6 \\
20:00:05.96 & 40:36:36.4 & 18437 &  41 & -21.9 \\
20:00:07.67 & 40:46:02.9 & 17868 & 115 & -19.6 \\
20:00:17.42 & 40:51:16.9 & 19187\tablenotemark{b} &  50 & -20.6 \\
\cutinhead{New Non-Cluster Velocities\tablenotemark{c}}
19:57:21.93 & 41:03:38.0 & 24763 &  99 & 18.3 \\
19:57:30.43 & 40:54:44.1 & 10373 &  72 & 20.4 \\
19:57:38.38 & 41:01:01.0 & 41162 & 105 & 18.9 \\
19:57:59.78 & 40:56:59.7 & 45209 &  80 & 20.4 \\
19:57:59.94 & 40:49:01.0 & 38764 & 102 & 20.0 \\
19:58:08.78 & 41:06:34.1 & 31748 &  90 & 18.2 \\
19:58:09.35 & 40:49:36.0 & 31419 &  82 & 19.4 \\
19:58:22.39 & 41:01:44.2 & 24613 &  77 & 18.8 \\
19:58:22.85 & 40:58:54.1 &  5966 &  77 & 18.3 \\
19:58:34.89 & 40:55:32.4 & 32318 &  76 & 18.4 \\
19:58:44.73 & 40:34:41.7 & 104540 & 104 & 19.8 \\
19:58:46.53 & 41:05:19.5 & 41792 &  70 & 19.7 \\
19:58:53.89 & 40:33:19.5 & 31299 &  88 & 19.8 \\
19:58:57.25 & 40:38:37.3 & 36006 &  82 & 18.8 \\
19:58:57.74 & 40:59:22.0 & 26082 &  82 & 18.9 \\
19:58:58.35 & 40:59:39.5 & 35796 &  66 & 19.3 \\
19:59:04.29 & 40:49:24.8 & 68294 &  93 & 18.2 \\
19:59:23.50 & 40:45:45.7 & 40592 & 111 & 19.1 \\
20:00:38.42 & 40:59:06.7 & 43830 & 116 & 19.9 \\
20:00:41.40 & 41:00:52.4 & 13431 &  94 & 19.7 \\
20:00:50.58 & 41:12:14.8 & 48897 &  75 & 20.1 \\
\enddata

\tablenotetext{a}{Emission-line redshift.}
\tablenotetext{b}{The velocity for this galaxy was incorrect in Paper 1.}
\tablenotetext{c}{Magnitudes for non-cluster galaxies are apparent magnitudes calculated for a 5\arcsec{} diameter aperture, and are not corrected for Galactic extinction.}

\end{deluxetable}

\begin{deluxetable}{r r r}
\tablecolumns{3}
\tablecaption{Missed Galaxies\label{tbl-novels}}
\tablewidth{0pt}
\tablehead{
\colhead{RA} & \colhead{Dec} & \colhead{$m_R$} \\
\colhead{(J2000)} & \colhead{(J2000)} & \colhead{(5\arcsec)}
}
\startdata
19:56:48.41 & 40:27:52.9 & 18.1 \\
19:56:48.99 & 40:56:40.6 & 18.1 \\
19:56:49.27 & 40:51:04.6 & 19.0 \\
19:56:49.28 & 40:48:48.6 & 19.4 \\
19:56:51.47 & 40:34:01.2 & 19.3 \\
19:56:52.03 & 40:36:19.0 & 19.0 \\
19:56:52.57 & 41:07:43.7 & 15.8 \\
19:56:53.77 & 40:27:53.2 & 19.6 \\
19:56:55.56 & 40:59:23.2 & 17.6 \\
19:56:55.82 & 41:03:51.3 & 18.4 \\
19:56:58.25 & 40:20:38.5 & 19.8 \\
19:57:00.04 & 40:50:02.7 & 17.9 \\
19:57:00.74 & 41:06:17.8 & 20.3 \\
19:57:00.75 & 40:48:43.6 & 18.7 \\
19:57:03.27 & 40:40:18.2 & 17.0 \\
19:57:04.00 & 40:17:04.0 & 18.7 \\
19:57:04.72 & 40:14:38.1 & 19.5 \\
19:57:04.73 & 41:00:43.2 & 19.9 \\
19:57:05.45 & 40:38:30.0 & 18.5 \\
19:57:05.75 & 40:15:10.4 & 17.5 \\
19:57:06.06 & 41:07:22.3 & 20.7 \\
19:57:06.12 & 40:49:53.8 & 18.3 \\
19:57:06.14 & 40:57:09.6 & 17.6 \\
19:57:07.12 & 40:54:41.8 & 19.3 \\
19:57:08.16 & 40:56:11.6 & 18.8 \\
19:57:08.61 & 41:11:40.4 & 18.9 \\
19:57:08.61 & 41:08:00.8 & 19.0 \\
19:57:09.11 & 40:31:13.6 & 20.0 \\
19:57:09.96 & 40:52:10.1 & 16.4 \\
19:57:10.62 & 41:01:21.6 & 20.8 \\
19:57:10.95 & 40:38:37.0 & 20.5 \\
19:57:11.72 & 41:08:15.2 & 18.5 \\
19:57:12.71 & 40:38:38.4 & 20.1 \\
19:57:16.18 & 40:39:13.5 & 20.1 \\
19:57:25.40 & 40:35:19.2 & 18.8 \\
19:57:26.37 & 41:02:01.7 & 20.2 \\
19:57:27.03 & 41:04:25.5 & 19.7 \\
19:57:28.11 & 40:56:26.1 & 21.2 \\
19:57:28.45 & 40:55:00.0 & 20.7 \\
19:57:30.19 & 40:51:16.4 & 20.7 \\
19:57:30.72 & 40:39:00.2 & 20.3 \\
19:57:32.29 & 40:14:33.2 & 19.4 \\
19:57:32.38 & 40:21:20.0 & 17.4 \\
19:57:33.30 & 40:40:48.7 & 19.9 \\
19:57:33.42 & 40:30:51.4 & 20.6 \\
19:57:33.75 & 41:13:02.7 & 19.8 \\
19:57:37.40 & 40:54:05.2 & 15.3 \\
19:57:37.96 & 41:09:26.8 & 18.4 \\
19:57:37.97 & 40:15:31.6 & 20.5 \\
19:57:39.52 & 40:46:39.9 & 20.2 \\
19:57:39.67 & 40:36:10.8 & 20.1 \\
19:57:39.82 & 41:13:02.1 & 19.9 \\
19:57:40.61 & 41:08:46.1 & 18.8 \\
19:57:41.86 & 41:10:18.4 & 18.6 \\
19:57:42.48 & 40:20:06.9 & 18.2 \\
19:57:43.72 & 41:03:27.3 & 18.7 \\
19:57:43.75 & 40:19:11.8 & 16.7 \\
19:57:44.16 & 41:02:11.5 & 19.9 \\
19:57:44.93 & 40:18:09.3 & 19.8 \\
19:57:45.95 & 40:21:53.2 & 20.2 \\
19:57:46.25 & 40:34:03.4 & 18.9 \\
19:57:46.70 & 40:53:12.1 & 20.3 \\
19:57:47.94 & 40:54:38.1 & 20.7 \\
19:57:48.57 & 40:41:00.5 & 19.2 \\
19:57:48.73 & 41:13:02.9 & 17.2 \\
19:57:49.52 & 40:33:39.1 & 20.9 \\
19:57:49.59 & 40:22:33.7 & 17.2 \\
19:57:49.67 & 40:36:17.3 & 18.7 \\
19:57:49.94 & 41:04:54.3 & 20.1 \\
19:57:50.38 & 40:25:37.1 & 18.8 \\
19:57:51.15 & 40:38:12.2 & 18.9 \\
19:57:53.82 & 40:55:10.6 & 20.9 \\
19:57:53.99 & 40:14:53.0 & 19.9 \\
19:57:55.18 & 40:57:14.5 & 19.6 \\
19:57:55.70 & 40:13:39.7 & 18.9 \\
19:57:57.06 & 40:42:07.6 & 20.1 \\
19:57:57.26 & 41:00:36.1 & 19.1 \\
19:57:57.59 & 40:56:17.3 & 18.0 \\
19:57:59.45 & 40:57:05.5 & 18.8 \\
19:58:00.83 & 40:56:54.9 & 19.5 \\
19:58:01.10 & 40:52:51.9 & 19.0 \\
19:58:01.84 & 40:38:45.7 & 17.9 \\
19:58:02.25 & 40:25:04.6 & 18.0 \\
19:58:02.25 & 40:14:48.6 & 18.4 \\
19:58:03.34 & 41:06:07.6 & 20.4 \\
19:58:04.07 & 40:19:43.1 & 20.3 \\
19:58:04.97 & 40:20:11.0 & 20.3 \\
19:58:04.99 & 40:55:09.7 & 17.0 \\
19:58:05.01 & 40:37:24.4 & 18.9 \\
19:58:05.59 & 40:37:01.9 & 18.7 \\
19:58:05.70 & 40:48:52.9 & 20.6 \\
19:58:06.62 & 40:55:44.8 & 19.3 \\
19:58:07.92 & 40:38:23.9 & 19.9 \\
19:58:08.38 & 41:05:14.2 & 19.6 \\
19:58:08.40 & 40:54:55.9 & 18.7 \\
19:58:08.73 & 40:57:45.4 & 19.3 \\
19:58:09.38 & 40:49:35.4 & 19.4 \\
19:58:09.98 & 41:12:54.0 & 19.0 \\
19:58:10.92 & 40:52:30.4 & 20.0 \\
19:58:13.98 & 40:24:38.4 & 20.7 \\
19:58:14.01 & 40:54:44.0 & 19.9 \\
19:58:14.77 & 40:26:58.6 & 19.6 \\
19:58:16.27 & 41:11:59.7 & 19.9 \\
19:58:16.55 & 41:00:39.4 & 20.0 \\
19:58:16.66 & 40:56:32.9 & 19.0 \\
19:58:18.88 & 40:45:22.1 & 20.3 \\
19:58:19.19 & 40:43:21.6 & 18.3 \\
19:58:19.25 & 40:40:02.9 & 17.8 \\
19:58:20.32 & 40:45:22.1 & 20.8 \\
19:58:20.71 & 40:57:30.2 & 20.0 \\
19:58:21.62 & 40:54:38.7 & 20.8 \\
19:58:21.96 & 41:09:49.9 & 19.0 \\
19:58:22.41 & 40:57:44.7 & 19.8 \\
19:58:22.48 & 41:01:42.6 & 18.8 \\
19:58:25.57 & 40:43:30.3 & 19.8 \\
19:58:26.40 & 41:02:54.7 & 20.2 \\
19:58:26.78 & 41:00:26.4 & 20.7 \\
19:58:26.84 & 40:58:47.1 & 19.9 \\
19:58:27.01 & 41:13:30.9 & 18.5 \\
19:58:27.35 & 40:37:08.0 & 17.9 \\
19:58:28.23 & 40:32:39.7 & 19.7 \\
19:58:28.85 & 40:40:04.1 & 17.7 \\
19:58:29.18 & 40:34:14.6 & 19.2 \\
19:58:30.05 & 41:07:59.0 & 19.8 \\
19:58:30.64 & 41:14:13.9 & 19.5 \\
19:58:32.49 & 41:05:16.7 & 17.8 \\
19:58:34.05 & 40:47:32.6 & 20.4 \\
19:58:36.32 & 40:44:58.3 & 19.7 \\
19:58:38.61 & 41:10:49.1 & 20.2 \\
19:58:39.21 & 40:58:43.1 & 20.5 \\
19:58:39.47 & 40:57:08.0 & 19.2 \\
19:58:39.76 & 40:57:32.2 & 18.6 \\
19:58:41.46 & 40:26:08.0 & 20.6 \\
19:58:41.61 & 40:56:56.3 & 19.4 \\
19:58:42.81 & 40:22:17.9 & 18.7 \\
19:58:42.90 & 40:45:46.1 & 20.0 \\
19:58:43.76 & 40:20:08.2 & 18.7 \\
19:58:43.99 & 41:06:14.8 & 21.1 \\
19:58:47.41 & 40:34:04.4 & 17.8 \\
19:58:48.43 & 41:02:25.5 & 17.6 \\
19:58:48.48 & 40:16:32.6 & 18.5 \\
19:58:48.64 & 40:16:13.2 & 17.6 \\
19:58:48.66 & 40:38:53.9 & 18.7 \\
19:58:49.84 & 40:25:41.4 & 20.9 \\
19:58:50.48 & 40:52:41.7 & 17.1 \\
19:58:51.28 & 40:45:48.1 & 19.7 \\
19:58:51.47 & 40:27:12.5 & 17.9 \\
19:58:51.64 & 40:50:17.8 & 18.3 \\
19:58:53.22 & 40:56:16.8 & 19.5 \\
19:58:53.43 & 40:59:37.4 & 19.2 \\
19:58:53.92 & 40:33:19.0 & 19.8 \\
19:58:53.94 & 40:44:34.1 & 19.9 \\
19:58:56.09 & 41:12:50.4 & 18.7 \\
19:58:56.31 & 40:51:58.8 & 19.7 \\
19:58:57.00 & 40:29:53.1 & 16.5 \\
19:58:58.08 & 40:13:42.3 & 17.7 \\
19:58:58.68 & 40:41:30.1 & 20.5 \\
19:58:59.57 & 41:07:22.3 & 20.9 \\
19:59:01.38 & 41:11:18.1 & 19.6 \\
19:59:03.46 & 40:56:08.3 & 20.2 \\
19:59:03.78 & 40:22:22.2 & 17.8 \\
19:59:04.23 & 40:46:56.6 & 17.8 \\
19:59:04.55 & 40:51:30.6 & 19.5 \\
19:59:04.88 & 40:19:07.1 & 19.9 \\
19:59:05.26 & 40:19:18.8 & 17.1 \\
19:59:05.86 & 40:54:04.4 & 20.8 \\
19:59:06.03 & 40:49:22.8 & 19.9 \\
19:59:08.69 & 40:54:35.3 & 19.7 \\
19:59:11.68 & 40:52:03.9 & 20.1 \\
19:59:12.42 & 40:18:40.2 & 18.2 \\
19:59:12.46 & 41:05:01.2 & 19.7 \\
19:59:13.50 & 40:16:05.6 & 20.3 \\
19:59:14.62 & 40:46:44.9 & 19.5 \\
19:59:15.21 & 40:54:43.8 & 19.1 \\
19:59:15.78 & 41:08:55.4 & 19.5 \\
19:59:16.32 & 41:04:36.3 & 21.0 \\
19:59:17.51 & 40:43:58.9 & 17.8 \\
19:59:17.93 & 40:22:00.6 & 20.1 \\
19:59:20.31 & 40:32:01.1 & 19.0 \\
19:59:21.68 & 40:27:01.1 & 20.4 \\
19:59:22.09 & 40:44:23.7 & 17.2 \\
19:59:22.24 & 40:25:54.4 & 19.1 \\
19:59:22.96 & 40:49:51.0 & 19.6 \\
19:59:23.89 & 40:51:43.3 & 18.9 \\
19:59:24.09 & 40:50:29.9 & 19.1 \\
19:59:24.21 & 40:49:19.9 & 19.5 \\
19:59:25.64 & 40:42:44.5 & 17.5 \\
19:59:26.24 & 40:44:48.2 & 19.4 \\
19:59:26.67 & 40:43:37.1 & 18.0 \\
19:59:27.15 & 40:14:40.2 & 19.7 \\
19:59:28.07 & 40:44:40.8 & 18.6 \\
19:59:28.52 & 40:21:07.1 & 18.5 \\
19:59:29.01 & 40:14:13.4 & 17.9 \\
19:59:31.22 & 40:24:16.6 & 17.5 \\
19:59:31.30 & 40:54:39.7 & 20.9 \\
19:59:32.07 & 40:14:30.8 & 19.4 \\
19:59:33.70 & 40:24:50.2 & 20.5 \\
19:59:34.84 & 40:39:42.8 & 18.0 \\
19:59:35.33 & 41:10:40.8 & 19.5 \\
19:59:35.66 & 40:45:32.5 & 17.9 \\
19:59:35.90 & 40:34:27.1 & 19.9 \\
19:59:36.66 & 40:56:42.2 & 20.6 \\
19:59:37.08 & 40:51:20.5 & 19.9 \\
19:59:37.15 & 40:37:59.3 & 20.6 \\
19:59:37.91 & 40:46:03.5 & 16.9 \\
19:59:38.15 & 40:47:58.1 & 19.7 \\
19:59:38.55 & 40:19:41.5 & 18.6 \\
19:59:39.85 & 40:47:21.3 & 18.1 \\
19:59:43.52 & 40:46:35.0 & 17.4 \\
19:59:43.63 & 40:25:41.2 & 17.8 \\
19:59:44.99 & 41:12:52.4 & 18.1 \\
19:59:45.00 & 40:18:22.8 & 17.7 \\
19:59:45.63 & 40:40:24.8 & 17.2 \\
19:59:47.36 & 40:23:32.2 & 17.6 \\
19:59:47.44 & 40:46:37.4 & 20.2 \\
19:59:48.49 & 40:58:01.9 & 17.6 \\
19:59:48.91 & 40:43:26.7 & 20.3 \\
19:59:49.07 & 40:32:18.4 & 18.6 \\
19:59:49.70 & 40:55:52.5 & 19.3 \\
19:59:49.85 & 40:56:22.4 & 20.2 \\
19:59:50.47 & 40:55:14.1 & 17.5 \\
19:59:51.80 & 40:53:35.3 & 19.0 \\
19:59:51.82 & 41:05:51.9 & 18.1 \\
19:59:52.28 & 40:41:36.9 & 19.8 \\
19:59:52.46 & 40:49:17.0 & 18.1 \\
19:59:52.89 & 40:44:17.7 & 19.0 \\
19:59:53.90 & 41:01:25.8 & 20.5 \\
19:59:54.01 & 40:24:13.2 & 18.3 \\
19:59:55.48 & 40:28:34.6 & 19.6 \\
19:59:55.56 & 40:23:17.2 & 19.1 \\
19:59:56.18 & 40:48:49.4 & 20.7 \\
19:59:56.24 & 40:26:22.9 & 17.3 \\
19:59:56.45 & 40:24:55.6 & 20.1 \\
19:59:56.88 & 40:30:55.7 & 18.2 \\
19:59:58.28 & 40:52:57.6 & 18.2 \\
19:59:59.77 & 40:22:04.8 & 18.7 \\
20:00:00.62 & 40:46:07.5 & 18.3 \\
20:00:00.75 & 40:43:17.0 & 19.2 \\
20:00:01.28 & 40:44:57.4 & 18.2 \\
20:00:02.04 & 40:39:52.6 & 19.9 \\
20:00:02.36 & 40:57:07.6 & 20.0 \\
20:00:02.62 & 40:57:06.1 & 20.1 \\
20:00:03.24 & 40:40:46.6 & 17.2 \\
20:00:03.51 & 40:24:01.3 & 19.9 \\
20:00:04.24 & 40:40:13.5 & 19.7 \\
20:00:04.25 & 40:57:27.7 & 18.7 \\
20:00:04.54 & 40:41:31.8 & 20.9 \\
20:00:06.74 & 41:07:53.6 & 18.5 \\
20:00:08.54 & 40:59:52.9 & 20.2 \\
20:00:10.06 & 41:08:12.6 & 20.8 \\
20:00:11.48 & 40:51:02.2 & 20.2 \\
20:00:12.60 & 40:54:18.4 & 19.3 \\
20:00:12.99 & 40:17:18.7 & 19.7 \\
20:00:13.68 & 40:54:32.1 & 20.5 \\
20:00:15.08 & 40:43:32.5 & 19.8 \\
20:00:15.50 & 41:10:00.0 & 19.8 \\
20:00:15.75 & 40:39:57.0 & 18.0 \\
20:00:18.03 & 40:55:34.2 & 20.8 \\
20:00:18.65 & 40:24:02.0 & 19.2 \\
20:00:19.69 & 40:54:10.1 & 19.3 \\
20:00:20.24 & 40:37:12.0 & 18.7 \\
20:00:20.31 & 40:51:46.3 & 17.3 \\
20:00:20.58 & 40:26:45.1 & 18.7 \\
20:00:21.94 & 40:53:57.3 & 20.8 \\
20:00:23.32 & 40:31:25.3 & 19.9 \\
20:00:23.64 & 40:45:52.1 & 20.4 \\
20:00:23.69 & 40:31:21.4 & 19.6 \\
20:00:23.91 & 40:32:48.8 & 19.5 \\
20:00:24.28 & 41:08:15.0 & 20.4 \\
20:00:24.37 & 40:56:26.9 & 19.7 \\
20:00:25.24 & 40:38:26.4 & 20.6 \\
20:00:27.57 & 40:40:03.1 & 20.2 \\
20:00:27.63 & 41:02:18.7 & 20.2 \\
20:00:27.82 & 40:27:55.9 & 18.7 \\
20:00:28.02 & 41:02:37.1 & 19.7 \\
20:00:29.86 & 40:25:51.4 & 20.0 \\
20:00:29.99 & 41:05:54.7 & 18.8 \\
20:00:30.30 & 40:46:24.6 & 17.1 \\
20:00:31.76 & 40:51:31.6 & 20.3 \\
20:00:32.87 & 40:37:46.6 & 19.9 \\
20:00:33.39 & 40:52:42.7 & 18.6 \\
20:00:33.62 & 40:34:35.6 & 19.2 \\
20:00:34.51 & 40:55:49.6 & 19.8 \\
20:00:34.83 & 40:34:48.8 & 17.7 \\
20:00:34.84 & 40:52:20.7 & 18.0 \\
20:00:37.51 & 40:46:45.8 & 20.1 \\
20:00:39.16 & 40:50:30.4 & 18.5 \\
20:00:39.20 & 41:02:38.4 & 20.2 \\
20:00:39.51 & 40:25:43.0 & 19.6 \\
20:00:40.81 & 40:57:18.7 & 18.5 \\
20:00:40.86 & 40:20:05.3 & 19.2 \\
20:00:40.92 & 40:40:48.6 & 17.6 \\
20:00:43.29 & 40:18:45.5 & 19.4 \\
20:00:43.47 & 40:30:50.6 & 21.1 \\
20:00:43.81 & 40:24:46.2 & 19.4 \\
20:00:45.03 & 40:51:31.2 & 20.6 \\
20:00:46.40 & 40:29:07.7 & 18.9 \\
20:00:47.43 & 40:39:55.5 & 17.6 \\
20:00:47.94 & 40:33:37.9 & 19.4 \\
20:00:49.02 & 40:53:40.3 & 19.6 \\
20:00:50.18 & 40:53:21.1 & 17.7 \\
20:00:51.39 & 40:28:38.7 & 18.0 \\
20:00:51.68 & 40:46:28.3 & 18.9 \\
20:00:51.97 & 40:32:56.1 & 19.2 \\
20:00:52.07 & 40:33:28.7 & 19.2 \\
20:00:52.51 & 41:01:36.0 & 19.2 \\
20:00:52.57 & 40:31:16.5 & 16.8 \\
20:00:52.67 & 40:50:32.7 & 18.0 \\
20:00:53.18 & 40:54:38.1 & 20.3 \\
20:00:53.96 & 40:58:57.2 & 19.6 \\
20:00:55.08 & 40:41:14.6 & 17.9 \\
20:00:55.21 & 40:44:30.5 & 18.8 \\
20:00:55.98 & 40:52:16.3 & 18.5 \\
20:00:56.02 & 41:01:54.5 & 18.5 \\
20:00:56.19 & 40:32:05.0 & 21.1 \\
20:00:56.75 & 40:54:01.7 & 18.9 \\
20:00:57.56 & 40:25:33.1 & 18.3 \\
20:00:59.44 & 40:23:18.9 & 19.5 \\
20:01:00.80 & 40:59:47.8 & 18.9 \\
20:01:01.44 & 40:47:55.3 & 20.1 \\
20:01:01.50 & 41:10:06.8 & 16.6 \\
20:01:03.32 & 41:00:54.0 & 18.8 \\
20:01:03.95 & 40:50:49.8 & 19.9 \\
20:01:04.68 & 40:40:19.9 & 18.8 \\
20:01:05.19 & 40:55:56.0 & 20.4 \\
20:01:06.35 & 40:56:35.0 & 19.7 \\
20:01:06.96 & 41:09:26.8 & 19.7 \\
20:01:07.21 & 40:50:47.8 & 17.9 \\
20:01:07.21 & 40:55:01.2 & 19.5 \\
20:01:07.30 & 40:50:16.4 & 18.7 \\
20:01:08.77 & 40:35:22.7 & 19.6 \\
20:01:08.99 & 40:32:07.0 & 20.2 \\
20:01:13.39 & 40:17:05.6 & 20.2 \\
20:01:14.79 & 40:33:10.9 & 19.5 \\
20:01:16.07 & 40:56:29.7 & 20.9 \\
20:01:18.09 & 40:54:48.6 & 19.9 \\
20:01:18.96 & 40:52:34.7 & 20.1 \\
20:01:19.99 & 40:41:43.0 & 20.3 \\
20:01:22.90 & 40:27:21.3 & 18.5 \\
20:01:23.76 & 40:26:01.8 & 17.2 \\
20:01:25.88 & 41:13:04.0 & 20.3 \\
20:01:29.64 & 40:53:22.4 & 19.1 \\
20:01:30.76 & 40:29:11.0 & 19.0 \\
20:01:31.03 & 40:50:44.8 & 19.4 \\
20:01:32.02 & 40:59:55.3 & 19.2 \\
20:01:33.20 & 40:37:31.6 & 18.2 \\
20:01:33.56 & 41:01:05.3 & 19.0 \\
20:01:33.57 & 40:45:14.4 & 17.6 \\
20:01:35.82 & 40:59:04.0 & 19.4 \\
20:01:38.28 & 40:31:55.1 & 18.8 \\
20:01:42.30 & 41:09:25.8 & 18.1 \\
20:01:42.69 & 40:38:42.1 & 19.9 \\
20:01:46.20 & 40:26:01.3 & 18.4 \\
20:01:46.75 & 40:36:00.3 & 19.5 \\
20:01:47.48 & 40:30:38.8 & 20.0 \\
20:01:47.74 & 41:11:08.2 & 20.1 \\
20:01:48.00 & 41:11:05.1 & 18.3 \\
20:01:48.31 & 40:35:57.6 & 20.2 \\
20:01:50.22 & 40:35:26.7 & 19.8 \\
20:01:51.72 & 40:25:22.5 & 17.4 \\
20:01:51.83 & 40:40:04.0 & 18.6 \\
20:02:04.97 & 41:07:32.6 & 18.5 \\
\enddata

\end{deluxetable}

\end{document}